# Leveraging Sample Entropy for Enhanced Volatility Measurement and Prediction in International Oil Price Returns

**Datta, Radhika Prosad**

Indian Institute of Foreign Trade (IIFT)

Kolkata Campus

1583 Madurdaha Chowbaga Road

Kolkata 700107

India

Email- rpdatta@gmail.com

*Abstract:*

*This paper explores the application of Sample Entropy (SampEn) as a sophisticated tool for quantifying and predicting volatility in international oil price returns. SampEn, known for its ability to capture underlying patterns and predict periods of heightened volatility, is compared with traditional measures like standard deviation. The study utilizes a comprehensive dataset spanning 27 years (1986-2023) and employs both time series regression and machine learning methods. Results indicate SampEn's efficacy in predicting traditional volatility measures, with machine learning algorithms outperforming standard regression techniques during financial crises. The findings underscore SampEn's potential as a valuable tool for risk assessment and decision-making in the realm of oil price investments.*

**Key Words**: *Sample Entropy, Volatility measurements, oil price returns, predictive modeling, Time series regression, machine learning, support vector machine (SVM), K nearest neighbours (KNN).*

**JEL Classification**: *: C22, G12, Q41, C53, C22,C45,C38,C45*

1. Introduction

Sample entropy (SampEn) is a complexity measure that has been used to assess the predictability of time series data. In the context of finance, SampEn has been employed as a measure of volatility, which is a crucial aspect of risk assessment in investment decisions. This paper aims to use SampEn in measuring volatility and use it to predict the traditional volatility measure standard deviation (std) in international Oil price returns.

SampEn as a Measure of Volatility quantifies the regularity and predictability of a time series by evaluating the conditional probability that patterns in the data remain similar over time. A higher SampEn value indicates greater regularity and predictability, while a lower value suggests higher irregularity and unpredictability. In the context of financial markets, a lower SampEn value is typically interpreted as higher volatility, implying a greater likelihood of unexpected price fluctuations.

Traditional volatility measures, such as standard deviation and variance, provide a simple assessment of the dispersion of a time series around its mean. However, they do not capture the underlying structure and patterns within the data. SampEn, on the other hand, offers a more nuanced understanding of volatility by taking into account the predictability of price movements. This makes SampEn a valuable tool for identifying periods of heightened volatility and potential risk in financial markets.

SampEn has been applied to various aspects of volatility analysis, including:

- Predicting market crashes: Studies have shown that SampEn can effectively predict periods of market crashes, as the regularity of price movements tends to decrease leading up to these events (Soloviev et al 2019), (Risso 2008)..

- Assessing stock market risk: SampEn has been used to assess the risk associated with individual stocks and portfolios (Cho 2016). A lower SampEn value for a particular stock suggests higher risk, indicating a greater likelihood of unexpected price fluctuations (Gao et al 2023).

- Analyzing commodity market volatility: SampEn has been employed to analyze the volatility of commodity markets, such as oil and gold (Benedetto et al 2015,). This information can be used by traders and investors to make informed trading decisions (Beneditto et al 2016).

Benefits of SampEn over Traditional Volatility Measures SampEn offers several advantages over traditional volatility measures:

- Captures underlying patterns: SampEn considers the patterns and regularity within a time series, providing a more comprehensive understanding of volatility.

- Predictive potential: SampEn has demonstrated the ability to predict periods of heightened volatility, making it a valuable tool for risk assessment.

- Adaptability to different time scales: SampEn can be applied to time series data of varying lengths, making it adaptable to different market conditions and trading strategies.

We apply the SampEn to quantify the volatility of Oil price returns by using a rolling window approach over a time period of 27 years from January 1986 to April 2023 and use it to predict the traditional volatility measure standard deviation (std) using standard time series regression and machine learning techniques. The results show that SampEn can predict the volatility of crude oil price returns as measured by the rolling std of oil price returns which is taken as the traditional measure of volatility. The time series based regression methods show that with ARIMA(4,1,3) model, the rolling standard deviation time series (ts_std) with the rolling SampEn time series (ts_SampEn) as an innovation vector is not able to capture the behaviour of ts_std completely since the residuals still have some autocorrelations as shown in the Results section. The machine learning techniques based on support vector machine SVM and K nearest neighbour (KNN) algorithms are able to predict the changes in the rolling windows std value during financial crisis in a better manner than Linear regression based methods.

The rest of this paper is organized as follows. We give a brief literature review in section 2 towards the end of which we point out the potential research gap and justify our research objective. Section 3 gives the methodology of calculating the SampEn. In Section 4 we give the details of our data and discuss the results of our calculations. Finally we conclude in section 5.

2. Literature Review

Sample entropy (SampEn) is a complexity measure that has been used to assess the predictability of time series data. In the context of finance, SampEn has been employed as a measure of volatility, which is a crucial aspect of risk assessment in investment decisions. This literature review aims to provide an overview of the application of SampEn in measuring volatility and its potential benefits over traditional volatility measures.

Some scientific papers that substantiate the literature review on sample entropy as a measure of volatility are given below:

In one of the first studies Bentes and Menezes (2012), discuss the potentialities of Shannon entropy and Tsallis entropy in understanding stock market volatility. The traditional measure, standard deviation, has drawbacks due to its sensitivity to extreme values. This paper explores the effectiveness of entropy, derived from physics, as an alternative measure of volatility compared to standard deviation.

Zhang et al (2019) propose a new method, rescaled range permutation entropy, emphasizing the ratio of range to standard deviation as a weight for time series segments to capture maximum volatility. Compared to other methods (sample entropy, fuzzy entropy, and weighted permutation entropy), it demonstrates superior discernibility in chaotic time series with extreme volatility. This suggests its potential as a valuable tool for assessing the complexity of nonlinear systems in such scenarios.

A paper by Ren et al (2021) mentions that Rescaled range permutation entropy, a recent method for assessing complexity in chaotic time series with extreme volatility, uses the ratio of range to standard deviation as a weight for vectors. While previous work has focused on typical chaotic systems, its intrinsic advancement and practical application potential remain unexplored. Simulations in this paper demonstrate its suitability for characterizing impulsive signals, and its significant discernibility in gas–liquid two-phase slug flow analysis suggests promising practical applications.

A paper by Sarlabous et al (2020) develops a sample entropy-based method to identify complex patient-ventilator interactions during mechanical ventilation. The authors demonstrate the ability of their method to detect changes in patient-ventilator interactions that are associated with clinical outcomes.

Namdhari and Li (2019) provides a comprehensive review of entropy measures, including sample entropy, and their application in quantifying uncertainty in stochastic processes. The authors discuss the advantages and limitations of different entropy measures and highlight the suitability of sample entropy for volatility analysis.

Sheraz et al (2015) investigate the use of various entropy measures, including sample entropy, to assess volatility in financial markets. The authors find that sample entropy provides a more effective measure of volatility than traditional measures, such as standard deviation and variance.

In a recent study Olbrys and Majewska (2022) compare the performance of approximate entropy (ApEn) and sample entropy (SampEn) in measuring volatility in financial time series data. The authors find that both ApEn and SampEn provide useful insights into volatility, but that SampEn is generally more robust and reliable.

The study by Chandra and Jhadao (2021) explores the feasibility of portfolio rotation strategies based on style, size, and time horizons using approximate entropy (ApEn) and sample entropy (SampEn) indicators computed from the India Volatility Index (India VIX). Comparing ApEn and SampEn with the change in India VIX, the study reveals that both entropy measures capture higher-order movements more effectively, indicating a superior indicator of volatile markets. SaEn particularly reflects fluctuations better. The findings suggest a computationally supported alternative for portfolio managers in the Indian market, offering potential benefits for enhancing returns and mitigating risks through strategic portfolio rotations.

Olbryś and Ostrowski (2021) introduced a novel methodology for measuring market depth using Shannon information entropy for high-frequency data, supported by an algorithm identifying trade initiators. The proposed entropy-based market depth indicator proves promising in directly measuring both market entropy and liquidity. Empirical experiments on real-time data from the Warsaw Stock Exchange confirm its effectiveness in comparing market depth and liquidity for different equities, providing a robust and statistically analyzed proxy for theoretical and empirical financial market analyses.

Taneja et al. (2019), December)- This paper compares various techniques for measuring volatility in the options market, including one based on the standard deviation, one based on the investor's expectation of future movements of the underlying asset price (implied), and one based on the concept of Shannon entropy. The empirical analysis is carried out to find some relationship between the three different approaches.

Ruiz et al (2012) - This paper proposes a new approach to measure volatility in energy markets using entropy. The authors apply the diffusion entropy method to the Nordic spot electricity market and estimate the scaling exponent.

Patra and Hiremath (2022) -This paper measures stock market efficiency by drawing a comprehensive sample from Asia, Europe, Africa, North-South America, and Pacific Ocean regions and ranks the cross-regional stock markets according to their level of informational efficiency.

Gradojevic and Caric, M. (2017)- This paper explores quantifying behavioral aspects of systemic risk using an entropy-based approach. Examining aggregate market expectations and systemic risk predictability during the 2008 financial crisis, it considers skewness premium and implied volatility ratio from out-of-the-money options. Results affirm the predictive and contemporaneous utility of the entropy setting in market risk management, with predictability linked to entropy type and underlying signal nature.

These are just a few examples of the many scientific papers that support the use of sample entropy as a measure of volatility. Sample entropy has been shown to be a valuable tool for understanding and predicting volatility in a wide range of financial markets and other complex systems.

Though many papers have used different information theoretic entropy related measures like the Shanon entropy, the Tsallis entropy, the Renyl entropy, the Approximate entropy ApEn and the Sample Entropy SampEn to study fluctuations in many financial and physiological time series data the application of the rolling windows SampEn (it has been shown by Olbrys and Majewska (2022) that SampEn is generally more robust and reliable) as a predictor for rolling windows standard deviation (std) has not been covered in any of the papers surveyed by the author.

3. Methodology

In this section we describe the methodology underlying the sample entropy calculations SampEn

The calculations reported in this paper, implement the algorithm code for SampEn written in R which was reported in the paper by Bonal (2019), therefore a comparable nomenclature has been employed.. Consider the following time sequence again v = {v(1), v(2), . . . , v( N )},a positive integer m, which satisfies 0≤ m ≤N, the length of the sequences to be compared denoted by N, and a real value r > 0, which indicates the tolerance level within which matches are admitted. For every computation, the variables N, m, and r must be fixed.

We then define the vectors
$y_m(i) = \{v(i), v(i + 1), \ldots, v(i + m − 1)\}$ (1)
and
$y_m(j) = \{v(j), v(j + 1), \ldots, v(j + m − 1)\}$. (2)
Then, the following formula is used to determine their Chebyshev distance:
$d[y_m(i), y_m(j)] = \max_{k=1,2,\ldots,m}(|v(i + k − 1) − v(j + k − 1)|)$. (3)

The next Equation (4) gives the maximum number of vectors $y_m(j)$ inside a radius r of $y_m(i)$ without allowing self-counting.

$C_j^m(r) = \frac{1}{(N-M-1)} \sum_{j=1, j \neq i}^{N-M}$ (number of times that $d[y_m(i), y_m(j)] \leq r$) (5)

The total number of potential vectors $B^m(r)$ is then determined using Equation (8) which shows the empirical likelihood of two sequences matching for m points.

$C_i^m(r) = \frac{1}{(N-M)} \sum_{i=1}^{N-M} C_i^m(r)$ (6)

In a similar manner, the next Equation (9) states the maximum number of vectors $y_{m+1}(j)$ at a distance r of $Y_{m+1}(i)$ without permitting self-matching:

$D_i^m(r) = \frac{1}{(N-M-1)} \sum_{j=1, j \neq i}^{N-M}$ (number of times that $d[y_{m+1}(i), y_{m+1}(j)] \leq r$) (7)

Next, we calculate the total number of matches $D^m(r)$ based on Equation (10). This denotes the empirical possibility that two sequences for m + 1 points (matches) are comparable.

$$D^m(r) = \frac{1}{(N-M-1)} \sum_{i=1,}^{N-M} D_i^m(r) \qquad (8)$$

Since It is obvious that the number of matching vectors ($D^m(r)$) is always less than or equal to the number of likely vectors ($C_j^m(r)$) that match. Therefore, the ratio ($D^m(r)/C^m(r)$) <1 is a conditional probability (Shannon 1948).

The time sequence v's SampEn value is calculated in the final step, as shown below:

$$SampEn(m, r, N)(v) = -\log(D^m(r)/C^m(r)) \qquad (9)$$

The SampEn(m, r, N) given by Equation (11) is the statistical estimator of the parameter SampEn(m, r):

$$SampEn(m, r) = \lim_{n \to \infty} [-\log(D^m(r)/C^m(r))] \qquad (10)$$

The Sample Entropy is close to zero for regular, repeating data points because the quantity ($D^m(r)/C^m(r)$) in Equation (9) approaches one (Richman & Moorman 2004).

4. Data and Results:

In this section we describe our data and results. The data consists of the price of crude oil as reported by the Federal Reserve Economic Data (FRED) and is available at https://fred.stlouisfed.org/series/DCOILWTICO. We consider the starting point of our data from 2nd January 1986 to 10th April 2023 a total of 9389 points. The following steps denote the details of our calculations:

1) In the first step we calculate the log returns of the prices of crude oil denoted by Rn given by the formula:

$$R_n = \ln \ln \left(\frac{P_n}{P_{n-1}}\right) \qquad (11)$$

Where Pn is the price of crude oil on day n and Pn-1 is the price of crude oil on day n-1 and ln is the natural logarithm.

2) We then calculate the standard deviation (std) and the sample entropy (SampEn) over rolling windows of size 252 data points which roughly translates into 1 year. After each calculation we shift the window by one step i.e. if the first window consists of data points 1 to 252, the second window consists of data points from 2 to 252. This is continued till the right hand side of the rolling window reaches the last data point on 10th April 2023. Our objective is to see how standard deviation and the sample entropy evolves with the passage of time.
3) Thus, we obtain the rolling std and SampEn for a total of 9137 points starting from 5th January 1987 to 10th February 2023 a total of 9137 points. Note that the first 252 points of our initial time series are required to calculate the std and SampEn for the first rolling window and so the std and SampEn calculations are available from the 253rd point onwards for the time series.
4) Our objective is to investigate if the sample entropy denoted by SampEn can be used to predict the standard deviation which is a standard measure of volatility of a time series in the case of the oil price time series.

Figures 1 to 2 show the time series plots for the standard deviation and sample entropy of our rolling windows calculations.

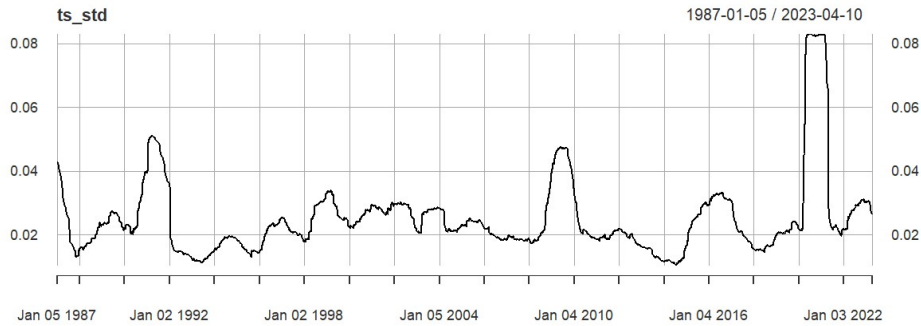

Figure 1 – Time series plot of standard deviations (ts_std) from the rolling windows calculations

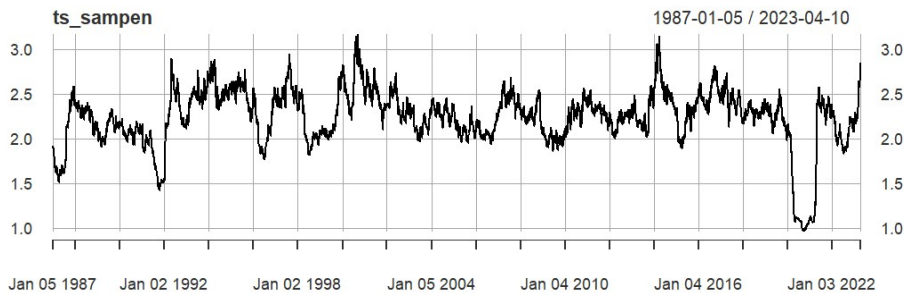

Figure 2 – Time series plot of sample entropies (ts_SampEn) from the rolling windows calculations

From the plots of the 2 time series we see that there is no obvious relation between them. The Pearson correlation coefficient which measures the strength and direction of a linear relationship between two variables is calculated to be -0.704 which shows that there is a significant negative correlation between the 2 time series.

We perform some tests on the 2-time series ts_std and ts_SampEn before we try to predict the standard deviation using the sample entropy.

1) An Augmented Dickey Fuller Test (ADF) is performed on ts_std with null hypothesis

   Ho: The time series is nonstationary and alternative hypothesis

   H1 : The time series is stationary

gives the following results:

ADF Statistic: -4.503642 , p value 0.01 which rejects the null hypothesis i.e. we conclude that the time series ts_std is stationary.

2) A similar Augmented Dickey Fuller Test (ADF) is performed on ts_SampEn with null hypothesis

   Ho: The time series is nonstationary and alternative hypothesis

   H1 : The time series is stationary

Gives the following results:

ADF Statistic: -4.942407, p value is 0.01 which again rejects the null hypothesis i.e. we conclude that the time series ts_SampEn is also stationary.

3) We now check for the presence of autocorrelations in the 2 time series ts_std and ts_SampEn. For this we use the Ljung Box test. The results of the Ljung Box test for ts_std gives a p-value < 2.2e-16 which implies that the null hypothesis can't be rejected i.e. the series is correlated.

Similarly, the results for the Ljung Box test for ts_SampEn gives p-value < 2.2e-16 which again implies that the null hypothesis can't be rejected i.e. the series is correlated.

So, we now have 2 time series ts_std and ts_SampEn that are both stationary but the individual time series are autocorrelated. Therefore, a simple linear regression model cannot be applied to predict one time series from the other as the basic statistical assumptions needed to perform simple linear regression is not met here.

Also, since the 2 series are individually stationary, looking at the cointegration between ts_std and ts_SampEn is also not feasible as the assumption for cointegration is that both the time series must be integrated of order 1 also denoted as I(1).

We are thus left with 2 choices:

a) Use a time series regression model.
b) Use machine learning techniques using a training and test set and evaluate the accuracy of the machine learning model through various error metrics like the mean squared error (MSE), the mean absolute error (MAE), the root mean squared error (RMSE) and the mean absolute percentage error (MAPE).

a) Using time series regression model:

We employ the auto.arima() function in R, as outlined by Hyndman (2018). This function is a refined version of the Hyndman-Khandakar algorithm (Hyndman & Khandakar, 2008), amalgamating unit root tests, AICc minimization, and Maximum Likelihood Estimation (MLE) to derive an ARIMA model. The regression is performed with $Y_t$ = ts_std as the dependent variable and $X_t$ = ts_SampEn as the predictor time series. The output of the regression with the best ARIMA model for the errors is obtained as:

Regression with ARIMA(4,1,3) errors

Coefficients:
```
         ar1      ar2     ar3      ar4      ma1     ma2      ma3
      0.3416  -0.2225  0.9202  -0.1101  -0.1351  0.1870  -0.8142
s.e.  0.0262   0.0218  0.0162   0.0137   0.0239  0.0206   0.0163
         xreg
      -0.0031
s.e.   0.0001
```

sigma^2 = 1.201e-07:  log likelihood = 59831.7
AIC=-119645.4   AICc=-119645.4   BIC=-119581.3

The fitted equation with the dependent variable (let's call it Y) and the predictor variable (xreg) can be written as follows:

$$Y_t = \phi 1 Y_{t-1,} + \phi 2 Y_{t-2,} + \phi 3 Y_{t-3,} + \phi 4 Y_{t-4,} + \theta 1 \varepsilon_{t-1} + \theta 2 \varepsilon_{t-2} + \theta 3 \varepsilon_{t-3} + \mu_t$$

where:

- $Y_t$ is the dependent variable at time $t$,
- $\phi 1, \phi 2, \phi 3, \phi 4$ are the autoregressive coefficients,
- $\theta 1, \theta 2, \theta 3$ are the moving average coefficients,
- $\varepsilon_t$ is the error term at time $t$,
- $\mu_t$ is the predictor variable (xreg) at time $t$.

Based on the provided coefficients, the equation becomes:

$$Y_t = 0.3416 Y_{t-1} - 0.2225 Y_{t-2} + 0.9202 Y_{t-3} - 0.1101 Y_{t-4} - 0.1351 \varepsilon_{t-1} + 0.1870 \varepsilon_{t-2} - 0.8142 \varepsilon t - 3 - 0.0031 \mu t$$

This equation represents the estimated relationship between the dependent variable $Y_t$, the predictor variable $X_t$, and the lagged dependent variables $Y_{t-1}, Y_{t-2},$ and $Y_{t-3,}$ in the context of our ARIMA(4,1,3) model.

We now need to check the residuals of fitting the above regression model to see if there are any remaining autocorrelations. The output from checking the residuals is given below:

Ljung-Box test

data: Residuals from Regression with ARIMA(4,1,3) errors
Q* = 35.161, df = 3, p-value = 1.127e-07

Model df: 7.   Total lags used: 10

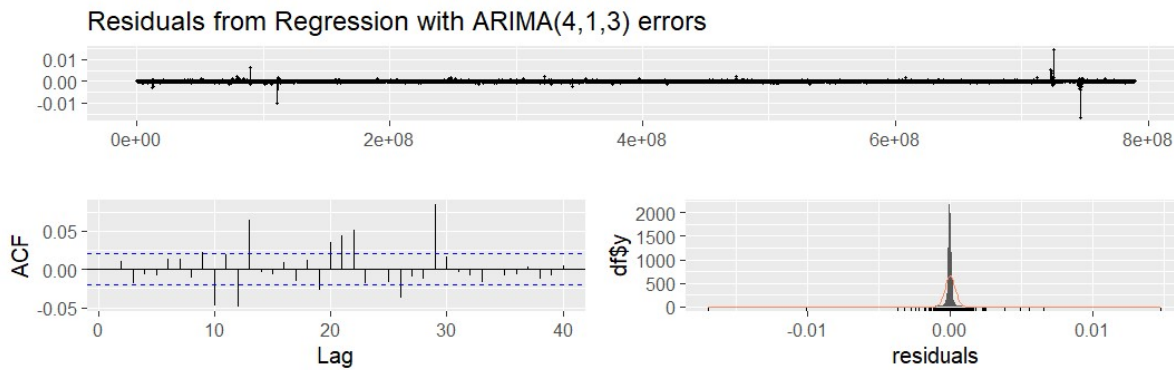

Figure 3 – Plot showing residuals from regression with ARIMA(4,1,3) errors, the ACF and the histogram of the residuals

The autocorrelation function (ACF) and the histogram of the residuals from the time series regression model of ts_std as the dependent variable and ts_SampEn as the predictor variable obtained by fitting the data show the following:

   i)      The time plot of the residuals shows some isolated peaks.
   ii)     The ACF shows some spikes at lags 10, 12 and 28 which are pronounced and go beyond the error levels. The Ljung Box test gives a p value of 1.127e-07. Since the p-value (1.127e-07) is less than the conventional significance level of 0.05, we reject the null hypothesis. This suggests that there is enough evidence to conclude that there is significant autocorrelation in the residuals at lags 1 to 9.
   iii)    The histogram of the residuals also differs significantly from the normal distribution.

iv) Thus overall, we may say that though some part of the rolling std series (ts_std) is indeed explained by our ARMIA(4,1,3) regression model with external regressor xreg= ts_SampEn i.e. the rolling sample entropy, there is still some unexplained part of ts_std which our model is not able to capture.

To generate forecasts using a regression model with ARIMA errors, it is essential to forecast both the regression and ARIMA components individually and then integrate the outcomes. Like standard regression models, predicting the regression part requires forecasting the predictors. When the predictors are readily available for future periods (e.g., calendar-related variables like time or day-of-week), this process is straightforward. However, if the predictors are unknown, two approaches can be employed: modelling them independently or relying on assumed future values for each predictor.

We will generate forecasts for the 300 days in the future using our fitted regression model by assuming that the future rolling windows std values will mirror the mean std observed over the past period which is considered in our calculations.

The results of this forecast are shown in the figure below:

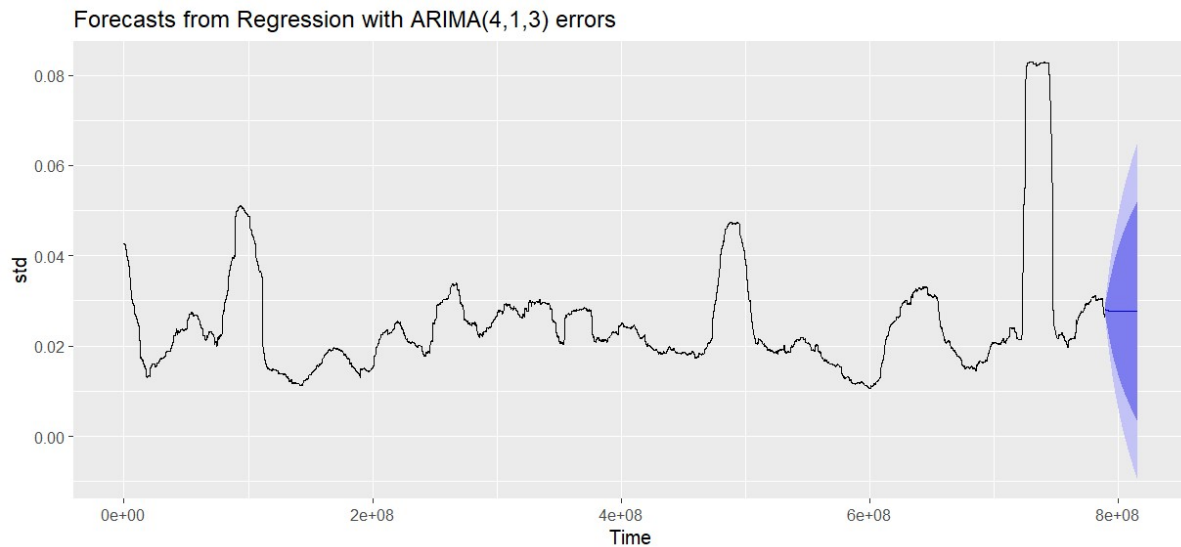

Figure 4: Forecasts from regression of ts_std as dependent variable and ts_SampEn as predictor variable.

As we can see from the figure the forecast for the rolling std using SampEn as the predictor variable is rather wide, the deep blue shaded region in the plot, the 95% confidence interval for the forecasts is shown using the light blue shaded region.

b) Using Machine learning techniques:

As it is well known, for prediction and forecasting purposes, machine learning methods are often preferred where it is difficult to model the time series by known statistical models. In this part of our research, we explore the use machine learning techniques to see if the time series of rolling window sample returns given by ts_SampEn can be used to forecast the time series of rolling window standard deviations denoted by ts_std

We use 3 simple methods here:

i) Simple linear regression with a training and test set ratio of 80:20

ii) Support vector machine or SVM regression with a training and test set ratio of 80:20
iii) K nearest neighbour or KNN regression with training and test set ratio of 80:20

Here the total number of data points is 9137, the training set consists of 7309 data points and the test set consists of 1828 data points.

|  | MAE | MAPE | MSE | RMSE |
|---|---|---|---|---|
| Linear Reg | 0.0069 | 29.3838 % | 7.5965e-05 | 0.00872 |
| SVM Reg | 0.0049 | 21.58587 % | 4.471e-05 | 0.006687 |
| KNN reg | 0.00511 | 23.24452 % | 4.65225e-05 | 0.00682 |

Table 1 - Errors from the machine learning calculations

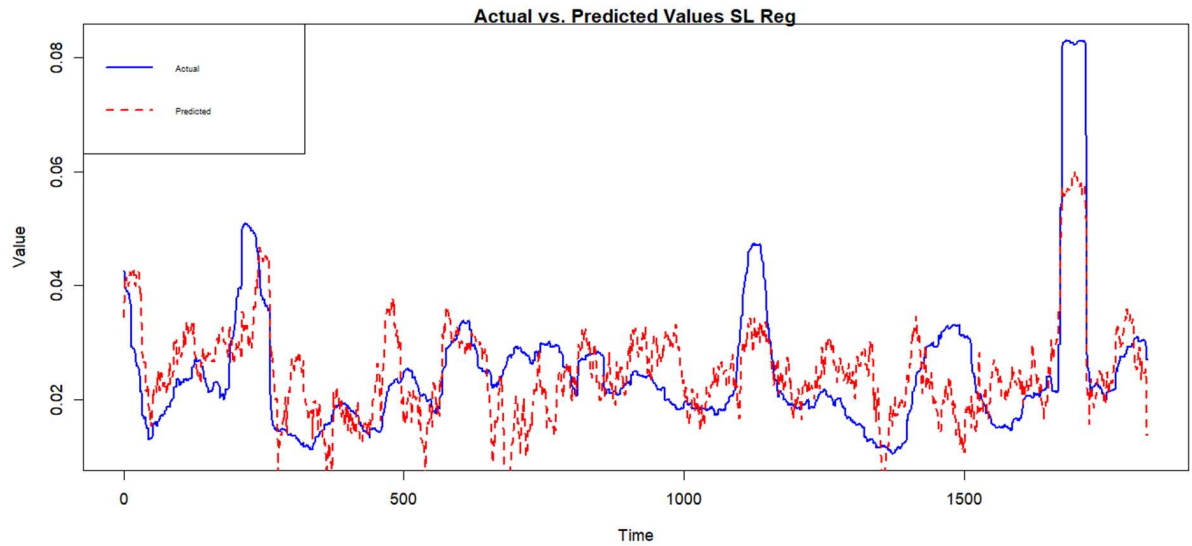

Figure – Actual versus predicted values of std on the test set using linear regression

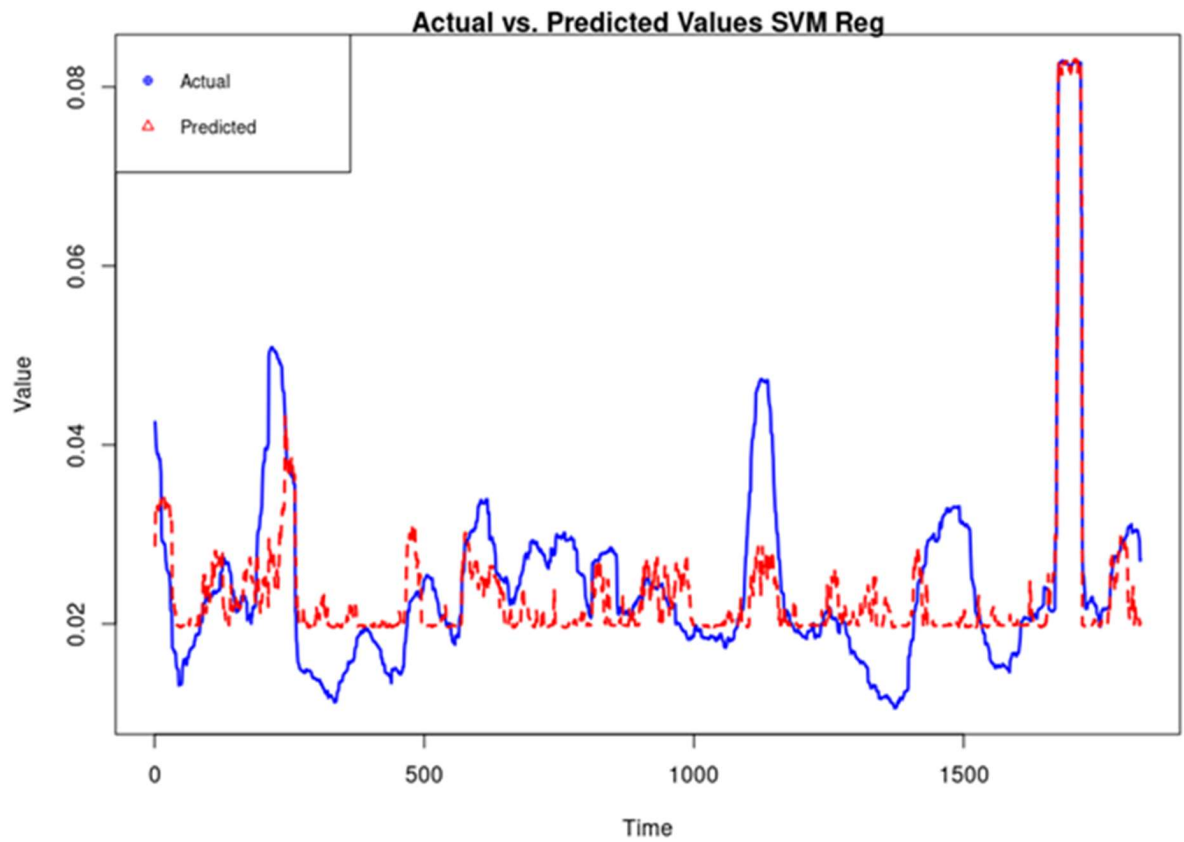

Figure – Actual versus predicted values of std on the test set using SVM regression

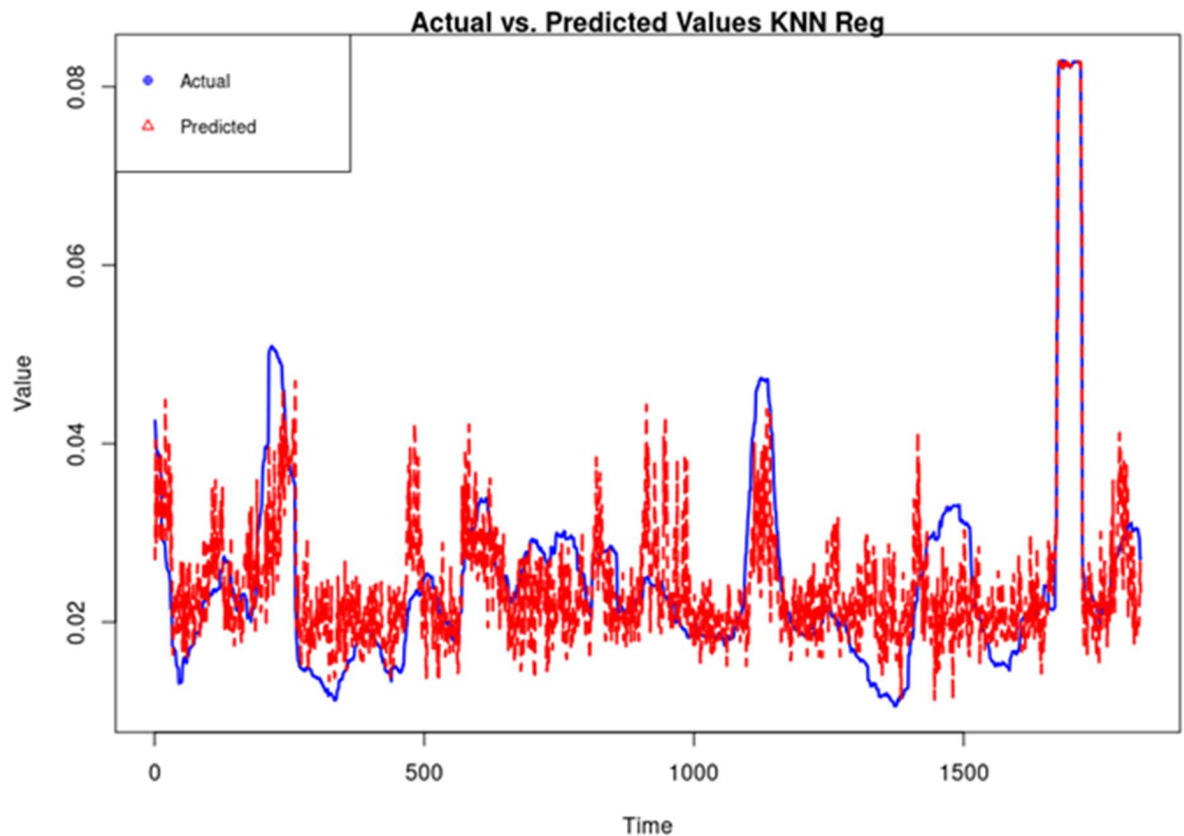

Figure – Actual versus predicted values of std on the test set using KNN regression

We can make the following observations from the results of Table 1.

a) Of all the machine learning methods used Linear Regression has the highest MAE, MAPE, MSE and RMSE values.
b) This is followed by the KNN regression method which has the next highest errors.
c) The SVM regression method gives the best values that are the lowest errors for all the error metrics.
d) The machine learning predictions using rolling SampEn are in general lower than the highest values in the actual rolling std values, similarly the predictions are higher than the actual lowest values. This means that the predictions in general exhibit lower fluctuations than the actual values.
e) If we look at the actual values of the calculated rolling std values we will observe that the values of the std start increasing sharply from 21st April 2020 when it has a value of 0.071265 and reach a maximum value of 0.08296 on 29th June 2020. Incidentally this period also coincides with the peak period of the global COVID 19 pandemic.
f) All the machine learning algorithms are able to predict this peak but again we observe that the linear regression based predictions are the worst, they predict the peak to be much lower.
g) The SVM and KNN algorithms predict the peak almost perfectly.
h) However though the SVM algorithm has the lowest values for the error metrics, the SVM algorithm consistently gives a higher value during the troughs (low values) and doesn't seem to predict the lower parts of the td_std series well.
i) The KNN algorithm shows the highest fluctuations in the predicted values over the actual values but the predictions follow the highs and lows of the actual values better than the SVM algorithm which seems to give better predictions during upswings rather than downswings in the actual ts_std time series values..

5. Conclusion:

In this paper we have used the information theoretic measure Sample Entropy denoted by SampEn to predict the rolling windows Standard Deviation over a period of 252 days for the daily returns of international crude oil prices from 2nd January 1986 to 10th April 2023 a total of 9389 points. The 252 days width of the rolling window is chosen to roughly approximate the duration of 1 year since there are approximately 252 trading days in a calendar year. We use both time series regression and machine learning methods (namely Linear Regression, Support vector machines SVM and K nearest neighbour KNN algorithms to predict the time series of rolling Standard Deviations using the time series of rolling SampEn as a predictor.

The results show that time series regression model with SampEn as the predictor time series (ts_ssampen) and ARIMA(4,1,3) errors is not able to completely explain the volatility in the oil price returns characterized by the rolling window Standard Deviation time series (ts_std) which is the target time series. Among the machine learning algorithms SVM has the least values for the error metrics but it fails to capture the troughs or downturns in the ts_std series, while the KNN algorithm in spite of having a slightly higher MAPE value 23.24% compared to SVM 21.59% is able to predict the troughs or downturns better than SVM. The Linear regression results show the least fluctuations among all the machine learning techniques. This can be attributed to the fact that Linear regression is intrinsically a linear technique while SVM and KNN are intrinsically nonlinear in nature. The relation between ts_std and ts_sampen is obviously nonlinear and SampEn which is a non parametric measure of the complexity of a time series is better able to capture the volatility which is apriori measured though the rolling window Standard Deviations (ts_std) for crude oil price returns.

Sample entropy has emerged as a powerful tool for measuring volatility and assessing risk in financial markets. Its ability to capture underlying patterns and predictability in time series data provides a more nuanced understanding of volatility compared to traditional measures. The predictive potential of SampEn further enhances its value in risk assessment and investment decision-making. As financial markets become increasingly complex and unpredictable, SampEn is likely to gain prominence as a valuable tool for volatility analysis and risk management.

References:

Benedetto, F., Giunta, G., & Mastroeni, L. (2015). A maximum entropy method to assess the predictability of financial and commodity prices. *Digital Signal Processing*, *46*, 19-31

Benedetto, F., Giunta, G., & Mastroeni, L. (2016). On the predictability of energy commodity markets by an entropy-based computational method. *Energy Economics*, *54*, 302-312.

Bentes, S. R., & Menezes, R. (2012, November). Entropy: A new measure of stock market volatility?. In *Journal of Physics: Conference Series* (Vol. 394, No. 1, p. 012033). IOP Publishing.

Chandra, A., & Jadhao, G. (2021). Profits Are in the Eyes of the Beholder: Entropy-Based Volatility Indicators and Portfolio Rotation Strategies. In *Computational Management: Applications of Computational Intelligence in Business Management* (pp. 69-95). Cham: Springer International Publishing.

Cho, J. K. (2016). Market Timing with the VKOSPI Sample Entropy Indicator. *International Journal of IT-based Business Strategy Management*, *2*(1), 17-24

Delgado-Bonal, A., & Marshak, A. (2019). Approximate entropy and sample entropy: A comprehensive tutorial. *Entropy*, *21*(6), 541.

Gao, Y. C., Tan, R., Fu, C. J., & Cai, S. M. (2023). Revealing stock market risk from information flow based on transfer entropy: The case of Chinese A-shares. *Physica A: Statistical Mechanics and its Applications*, 128982.

Gradojevic, N., & Caric, M. (2017). Predicting systemic risk with entropic indicators. *Journal of Forecasting*, *36*(1), 16-25.

Hyndman, R. J., & Athanasopoulos, G. (2018). Forecasting: principles and practice, OTexts: Melbourne, Australia. OTexts. com/fpp2.

Hyndman, R. J., & Khandakar, Y. (2008). Automatic time series forecasting: the forecast package for R. *Journal of statistical software*, *27*, 1-22.

Namdari, A., & Li, Z. (2019). A review of entropy measures for uncertainty quantification of stochastic processes. *Advances in Mechanical Engineering*, *11*(6), 1687814019857350.

Olbrys, J., & Majewska, E. (2022). Approximate entropy and sample entropy algorithms in financial time series analyses. *Procedia Computer Science*, *207*, 255-264.

Olbryś, J., & Ostrowski, K. (2021). An entropy-based approach to measurement of stock market depth. *Entropy*, *23*(5), 568

Patra, S., & Hiremath, G. S. (2022). An Entropy Approach to Measure the Dynamic Stock Market Efficiency. *Journal of Quantitative Economics*, *20*(2), 337-377.

Richman, J. S., Lake, D. E., & Moorman, J. R. (2004). Sample entropy. In *Methods in enzymology* (Vol. 384, pp. 172-184). Academic Press.

Risso, W. A. (2008). The informational efficiency and the financial crashes. *Research in international business and finance*, *22*(3), 396-408.

Ren, W., Zhang, J., & Jin, N. (2021). Rescaled range permutation entropy: a method for quantifying the dynamical complexity of gas–liquid two-phase slug flow. *Nonlinear Dynamics*, *104*(4), 4035-4043.

Ruiz, M. D. C., Guillamón, A., & Gabaldón, A. (2012). A new approach to measure volatility in energy markets. *Entropy*, *14*(1), 74-91.

Shannon, C. E. (1948). A mathematical theory of communication. *The Bell system technical journal*, *27*(3), 379-423.

Sheraz, M., Dedu, S., & Preda, V. (2015). Entropy measures for assessing volatile markets. *Procedia Economics and Finance*, *22*, 655-662.

Soloviev, V., Serdiuk, O., Semerikov, S., & Kohut-Ferens, O. (2019, October). Recurrence entropy and financial crashes. In *2019 7th International Conference on Modeling, Development and Strategic Management of Economic System (MDSMES 2019)* (pp. 385-388). Atlantis Press

Taneja, H. C., Batra, L., & Gaur, P. (2019, December). Entropy as a measure of implied volatility in options market. In *AIP Conference Proceedings* (Vol. 2183, No. 1, p. 110005). AIP Publishing LLC.

Zhang, J. C., Ren, W. K., & Jin, N. D. (2020). Rescaled range permutation entropy: a method for quantifying the dynamical complexity of extreme volatility in chaotic time series. *Chinese Physics Letters*, *37*(9), 090501.

.